\documentclass[preprint,3p,onecolumn]{elsarticle}
\pdfoutput=1

\usepackage{graphicx}
\usepackage{amsmath,amssymb,nccmath}
\usepackage{caption}
\usepackage{subcaption}
\usepackage{hyperref}
\usepackage{soul}
\hypersetup{colorlinks,linkcolor=blue}
\begin{document}%

\newcommand{\beq}{\begin{equation}}
\newcommand{\eeq}{\end{equation}}
\newcommand{\bea}{\begin{eqnarray}}
\newcommand{\eea}{\end{eqnarray}}
\newcommand{\bal}{\begin{aligned}}
\newcommand{\eal}{\end{aligned}}
\newcommand{\rf}[1]{(\ref{#1})}
\def\al{\alpha}
\def\be{\beta}
\def\ga{\gamma}
\def\de{\delta}
\def\ep{\epsilon}
\def\ze{\zeta}
\def\et{\eta}
\def\th{\theta}
\def\ka{\kappa}
\def\la{\lambda}
\def\rh{\rho}
\def\vr{\varrho}
\def\si{\sigma}
\def\vs{\varsigma}
\def\ta{\tau}
\def\up{\upsilon}
\def\ph{\phi}
\def\ch{\chi}
\def\ps{\psi}
\def\om{\omega}
\def\Ga{\Gamma}
\def\De{\Delta}
\def\Th{\Theta}
\def\La{\Lambda}
\def\Si{\Sigma}
\def\Up{\Upsilon}
\def\Ph{\Phi}
\def\Ps{\Psi}
\def\Om{\Omega}
\def\Gat{{\tilde \Ga}}
\def\cG{{\cal G}}
\def\cL{{\cal L}}
\def\cM{{\cal M}}
\def\cN{{\cal N}}
\def\cW{{\cal W}}
\def\cR{{\cal R}}
%
%repeating greek index combination abbreviations
\def\mn{{\mu\nu}}
\def\ab{{\al\be}}
\def\gd{{\ga\de}}
\def\bg{{\be\ga}}
\def\abgd{{\al\be\ga\de}}
\def\agbd{{\al\ga\be\de}}
\def\abg{{\al\be\ga}}
\def\bgd{{\be\ga\de}}
%%%%%

\def\prt{\partial}

\def\pt#1{\phantom{#1}}
\def\sb{\overline{s}}
\def\Ct{\tilde{C}}

\def\tg{{\tilde g}}
\def\tt{{\tilde t}}
\def\tR{{\tilde R}}
\def\tr{{\tilde r}}

\def\xb{\overline {x}}
\def\oprt{{\overline \partial}}
\def\oA{\overline {A}}
\def\oF{\overline {F}}
\def\oR{\overline {R}}
\def\ovr{\overline {r}}
\def\oh{\overline {h}}
\def\ot{\overline {t}}

%others
%\def\fr#1#2{{{#1} \over {#2}}}
%\def\half{{\textstyle{1\over 2}}}
%\def\quar{{\textstyle{1\over 4}}}
%\def\rf#1{(\ref{#1})}

\newcommand{\RX}[1]{\textcolor{cyan}{#1}}  % Rui Xu

\journal{Physics Letters B}

%\begin{document}

\begin{frontmatter}

\title{Classical radiation fields for scalar, electromagnetic, and gravitational waves with spacetime-symmetry breaking}

\author[inst1]{Quentin G. Bailey}
\author[inst1]{Alexander S. Gard}

%\affiliation[inst1]{organization=Embry-Riddle Aeronautical University,3700 Willow Creek Road, Prescott,86301,AZ,USA}
\address[inst1]{Embry-Riddle Aeronautical University, 3700 Willow Creek Road, Prescott, 86301, AZ, USA}

\author[inst2]{Nils A. Nilsson}

\address[inst2]{SYRTE, Observatoire de Paris, Universit\'e PSL, CNRS, Sorbonne Universit\'e, LNE, 61 avenue\\ de l'Observatoire, Paris, 75014, France}

% \affiliation[inst2]{organization={SYRTE, Observatoire de Paris, Universit\'e PSL, CNRS, Sorbonne Universit\'e, LNE},%Department and Organization
%             addressline={61 avenue de l'Observatoire}, 
%             city={Paris},
%             postcode={75014}, 
%             state={},
%             country={France}}
            
\author[inst3,inst4]{Rui Xu}
\author[inst4,inst5]{Lijing Shao}

\address[inst3]{Department of Astronomy, Tsinghua University, Beijing, 100084, China}

% \affiliation[inst3]{organization={Department of Astronomy},%Department and Organization
%             addressline={Tsinghua University}, 
%             city={Beijing},
%             postcode={100084}, 
%             state={},
%             country={China}}

\address[inst4]{Kavli Institute for Astronomy and Astrophysics, Peking University, Beijing, 100871, China}

% \affiliation[inst4]{organization={Kavli Institute for Astronomy and Astrophysics},%Department and Organization
%             addressline={Peking University}, 
%             city={Beijing},
%             postcode={100871}, 
%             state={},
%             country={China}}

\address[inst5]{National Astronomical Observatories, Chinese Academy of Sciences, Beijing, 100012, China}

% \affiliation[inst5]{organization={National Astronomical Observatories},%Department and Organization
%             addressline={Chinese Academy of Sciences}, 
%             city={Beijing},
%             postcode={100012}, 
%             state={},
%             country={China}}

\begin{abstract}
An effective field theory framework is used to investigate some Lorentz-violating effects on the generation of electromagnetic and gravitational waves, 
complementing previous work on propagation.
Specifically we find solutions to a modified, 
anisotropic wave equation, 
sourced by charge or fluid matter. 
We derive the radiation fields for scalars, 
classical electromagnetic radiation, 
and partial results for gravitational radiation.
For gravitational waves, 
the results show longitudinal and breathing polarizations proportional to coefficients for spacetime-symmetry breaking.
\end{abstract}

\end{frontmatter}

\section{Introduction}
\label{sec:intro}

Presently, 
interest in tests of foundations of General Relativity (GR) is high, 
including both theory and experiment.
Motivation for these studies include the possibility that some aspects of foundations of GR may be modified in a unified theory of physics that incorporates quantum gravity.
In particular, 
suggestions that spacetime-symmetry foundations of GR, 
like local Lorentz symmetry, 
could be broken in small but potentially detectable ways \cite{ksstring89,kp95} 
has motivated a plethora of theoretical studies and analyses 
\cite{Bailey_2023,Mariz:2022oib,Petrov:2020wgy,safronova18,Will:2014kxa,Tasson:2014dfa,Liberati13,datatables}.

Much theoretical work has been accomplished within effective-field theory (EFT) descriptions 
of spacetime-symmetry breaking, 
as well as with specific models.
This includes extensive literature
on the effects for electromagnetic waves and gravitational waves propagating in the vacuum \cite{Carroll:1989vb,km02,km09}.
Also, 
studies using non-EFT approaches abound in the literature \cite{Amelino-Camelia:2008aez,Mattingly:2005re}.
Accomplishments in Quantum Field Theory studies of spacetime-symmetry breaking are now prolific \cite{Jackiw:1999yp,Kostelecky:2000mm,Kostelecky:2001jc,Altschul:2019eip,Carvalho:2018vtr,Reyes:2014wna,Nascimento:2014owa,Reyes:2013nca,Cambiaso:2012vb,Casana:2010nd,Brito:2008ec,Colladay:2007aj}.
Much of the latter work relies on solutions 
to the field equations in momentum space, 
which is what is needed for QFT applications \cite{Casana:2009xs,Ferreira:2019ygi}.
Relatively few works have developed classical position-space solutions for the Green functions \cite{Casana:2008sd,bk04}, 
in particular, 
classical radiation multipole expansions
seem to be scant \cite{Reddy:2017smc,Amarilo:2018zqg}, 
in the EFT description of spacetime symmetry breaking.

The purpose of this article is to obtain general position-space solutions and study wave generation in the context of spacetime-symmetry breaking described by an EFT
\cite{kp95,ck97,ck98}.
Rather than a comprehensive study, 
we focus on minimal terms in the EFT
and use a coordinate transformation trick to find the exact Green function for a modified wave equation.
Our results are then applied to scalar fields, 
the electromagnetic sector, 
and the gravitational sector with some intriguing partial results 
on gravitational wave polarizations.
We also compare briefly to perturbative approaches.

Except when we discuss some results for gravitational waves, 
most of this work is in flat spacetime with the metric signature $-+++$, 
we use Greek letters for spacetime indices, 
and latin letters for spatial indices.
For the notation we follow conventions of other references on spacetime-symmetry breaking \cite{ck98,k04,bk06}.

\section{Green function with modified wave operator}
\label{Green function}

In effective-field theory descriptions of spacetime-symmetry breaking, 
one encounters Lagrange densities of the schematic form
$\cL \supset \et^\mn \prt_\mu \ps \prt_\nu \ps+ t^{\mu\nu\la...} \prt_\mu \psi \prt_\nu \prt_\la ... \ps $, 
for some field $\ps$,
where $t^{\mu\nu\la...}$ is a generic set of coefficients describing the degree of symmetry breaking for the field \cite{ck98,Edwards:2018lsn}.
Upon obtaining the field equations,
one typically encounters wave-type equations modified from 
the usual D'Alembertian operator $\Box = \prt^\al \prt_\al = \nabla^2 - \prt_t^2$;
to solve them, 
one can seek a Green function solution.

For actions with just two derivatives,
the typical problem involves finding a Green function $G(x,x^\prime )$ satisfying 
the equation
\beq
(\tilde g)^\mn \prt_\mu \prt_\nu G (x,x^\prime) = -\de^{(4)} (x-x^\prime),
\label{GreenEqn}
\eeq
where ${\tilde g}^\mn$ are constants.
These constants ${\tilde g}^\mn$ can be chosen so that there is a well-posed hyperbolic partial differential equation 
for the smooth source case, 
(i.e., for the underlying equation we are trying to solve $(\tilde g)^\mn \prt_\mu \prt_\nu \ps=\rh$) \cite{Wald:1984rg}.
Specifically, 
we will assume the following generic form:
\beq
\tg^\mn = \et^\mn + k^\mn,
\label{modMetric}
\eeq
where $k^\mn$ are a set of constant coefficients assumed to have values 
in the chosen coordinates sufficiently less than unity, 
so that $\tg^\mn$ is guaranteed an inverse.
Using Fourier transform methods, 
the momentum space solution of \rf{GreenEqn} is relatively trivial, 
while to date, 
no exact position space solution has been explicitly written and studied, 
although results can be found in certain limits \cite{bk04,Ferreira:2019ygi}.

The solution to \rf{GreenEqn} can be obtained by changing coordinates \cite{bk04,Altschul:2006zz}
%add cite
so that the equation appears with the conventional wave operator.
Specifically,
we change coordinates $x^\mu = x^\mu ( \xb^\nu ) $, 
in a particular way such that under this coordinate change,
\beq
{\overline \tg}^\mn = \frac {\prt \xb^\mu}{\prt x^\al}
\frac {\prt \xb^\nu}{\prt x^\be} \tg^{\al\be}= \et^\mn,
\label{gchange}
\eeq
so that ${\overline \tg}^\mn$ takes on the numerical values of the Minkowski metric. 
Such a transformation can generally be shown to exist with mild assumptions on $k^\mn$, 
for example, 
one can write such a transformation using a series $\xb^\mu = x^\mu - \frac 12 k^\mu_{\pt{\mu}\al} x^\al +...$.
Care is required here because the spacetime metric in the $\xb^\mu$ system is {\bf not} Minkowski.\footnote{This type of procedure was carried out at leading order in the appendix of
Ref.\ \cite{bk04} to demonstrate the physical equivalence of having certain forms of Lorentz violation
in the photon sector or the matter sector. Also see Ref.\ \cite{Altschul:2006zz}.}
In the new coordinate system, 
the equation \rf{GreenEqn} is
\beq
\et^\mn \oprt_\mu \oprt_\nu G(\xb,\xb^\prime ) = -\frac {1}{\sqrt{-\tg}}\de^{(4)} (\xb - \xb^\prime ),
\label{GreenEqn2}
\eeq
which resembles the standard wave operator Green function equation.
The determinant of $\tg^\mn$ is denoted $\tg$. 
Note that, 
despite appearances, 
one cannot generally remove $k_\mn$ from the framework altogether if there is a matter sector \cite{bk04,Altschul:2007kr,kt11,Bonder:2015maa}.
Only in the vacuum solution can one eliminate the coefficients $k_\mn$ entirely.

The solution to \rf{GreenEqn2} is a standard one up to a scaling \cite{Jackson}, 
$G = \de ( \et_\mn (\xb-\xb^\prime )^\mu (\xb-\xb^\prime )^\nu  )/ 4\pi \sqrt{-\tg}$.
One then transforms this function back to the original coordinate system:
\beq
\bal
G (x,x^\prime) &= \frac {1} {2\pi \sqrt{-\tg }} \de 
\left( -(\tg^{-1})_\mn (x-x^\prime )^\mu (x-x^\prime )^\nu \right),
\\
&= \frac {1} {4\pi \sqrt{-\tg} } 
\frac {1} { \tR }
\de (t^\prime - \tt_R ).
\label{GreenResult}
\eal
\eeq
In this expression
we use a modified retarded time $\tt_R$ and modified distance $\tR$:
\beq
\bal
\tt_R &= t - \frac { \tR +(\tg^{-1})_{0i}R^i}{(\tg^{-1})_{00}},
\\
\tR &=\sqrt{-(\tg^{-1})_{00} (\tg^{-1})_{ij} R^i R^j + ((\tg^{-1})_{0i} R^i)^2 },
\label{ttR}
\eal
\eeq
where
$R^i = (x-x^\prime)^i$.
The first line of \rf{GreenResult}
forces an evaluation along a skewed light cone $-(\tg^{-1})_\mn (x-x^\prime )^\mu (x-x^\prime )^\nu=0$.
The second line breaks up the delta function, 
and the choice of retarded boundary conditions is made.
This result will be used for the scalar, 
vector and tensor examples to follow. 

\section{Scalar Example}
\label{scalar example}

\subsection{Exact solution}
\label{exact solution}

We apply the results of the Green function \rf{GreenResult} 
to the case of a real scalar field with generic source function.
Thus we solve the equation
\beq
(\et^\mn + k^\mn ) \prt_\mu \prt_\nu \ps = -\rh,
\label{scalar}
\eeq
where $\rh$ stands for a generic source density for the scalar.
Using the general Green function results above, 
we obtain,
\beq
\ps = \frac {1}{4\pi \sqrt{-\tg} } 
\int d^3 r^\prime \frac {\rh (\tt_R , \vec r^\prime ) }{\tR}.
\label{psi}
\eeq

For calculations of ``wave zone" results, 
we will use an expansion similar to that
in Ref.\ \cite{pw14}, 
wherein the authors construct a systematic wave zone and near zone expansion.  
We start by assuming that the field point $\vec r$ is located far outside of the source region where $\rh \neq 0$;
thus the source $\rh$ has ``compact support".
If this is the case, 
then we may use a series expansion assuming $r \gg r^\prime$.
To expand the time argument of $\rh$, 
we must also assume the characteristic wavelength of $\ps$ is larger than the scale of the source $\la>r^\prime$.
It will be useful to use the following quantities, 
obtained by evaluating the expressions \rf{ttR} above when $\vec r^\prime =0$:
\beq
\tr = \sqrt{-(\tg^{-1})_{00} (\tg^{-1})_{ij} r^i r^j + ((\tg^{-1})_{0i} r^i)^2 },
\quad
\tt_r = t - \frac {\tr + (\tg^{-1})_{0i}r^i } { -(\tg^{-1})_{00} }.
\label{ttr}
\eeq
Following parallel steps to Ref.\ \cite{pw14} (Section 6.3), 
we arrive at the series:
\beq
\ps = \frac {1}{4\pi \sqrt{-\tg} }
\sum_{l=0}^{\infty} 
\frac {(-1)^l} {l!}
\prt_L \left( \frac {1}{\tr} 
\int d^3 r^\prime \rh (\tt_r , \vec r^\prime ) r^{\prime L} \right).
\label{psiWZ}
\eeq
In this expression we use the index abbreviation $L=i_1 i_2 i_3...i_l$.
For what follows we define a tangent vector $N_j = - \prt_j \tt_r$,
which reduces to the unit vector $n^j = r^j/r$ 
when $k_\mn \rightarrow 0$, 
and represents the direction of wave propagation.

It is useful to note some results that are leading order in the coefficients $k^\mn$.
Using the definition \rf{modMetric}, 
we have for the inverse metric,
modified retarded time $\tt_r$, 
and the tangent vector $N_j$,
respectively
\beq
\bal
(\tg^{-1})_\mn &= \et_\mn-k_\mn,\\
\tt_r &= t - r (1 - \frac 12 k_{00} - \frac 12 k_{ij} n^i n^j ) + k_{0i}r^i, \\
N_i &= n_i ( 1 - \frac 12 k_{00} + \frac 12 k_{jk} n^j n^k  )
-k_{ij} n^j
-k_{0i}.
\label{N1}
\eal
\eeq
to first order in $k_\mn$.
Using these approximations we obtain the first 3 terms of the series \rf{psiWZ} in the wave zone (keeping only terms with $1/r$ falloff):
\beq
\bal
\ps &= \frac {1}{4\pi r} \Big( 
Q [ 1- \tfrac 12 k_{00} + \tfrac 12 k_{ij}n^i n^j ]
+ \dot P^i [n_i (1- k_{00} + k_{jk}n^j n^k ) 
-k_{ij}n^j - k_{0i} ]
\\
&
\pt{space}
+\frac 12 \ddot I^{ij}
[n_i n_j ( 1- \tfrac 32 k_{00} + \tfrac 32 k_{lm}n^l n^m) -2 n_i k_{jk} n^k - 2 k_{0i}n_j]+... \Big)|_{t=\tt_r}.
\eal
\label{psiLO}
\eeq
Here $Q$ is the total ``charge", $\vec P$ is the dipole moment and 
$I^{ij}$ is the inertia tensor
associated with the source density $\rh$.
It is critical to note that the terms on the right-hand side of \rf{psiLO} are evaluated at the modified retarded time in \rf{ttr} and \rf{N1}.
This implies a deformed dependence on the space and time coordinates of the field point.

\begin{figure}
     \centering
     \begin{subfigure}[b]{0.4\textwidth}
        \centering
\includegraphics[scale=.35]{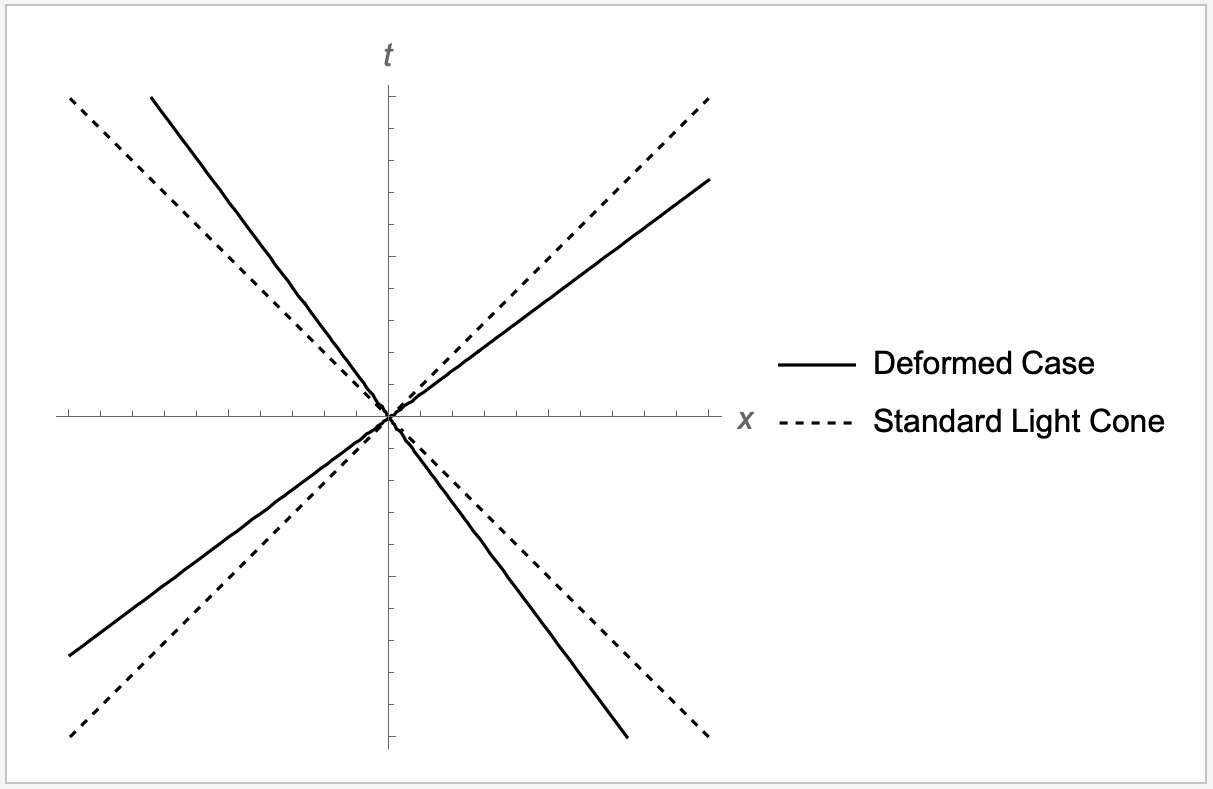}
\caption{A $t-x$ spacetime diagram illustrating the modified wave propagation when the coefficients $k_\mn$ are nonzero.  The solid curve is the solution of $t^2 - x^2 +2 k_{01} t x=0$, and the dashed curve is the usual light trajectory $t=\pm x$.
The size of the coefficient is exaggerated at $k_{01}=-0.3$.}
\label{fig1}
\end{subfigure}
     \hfill
     \begin{subfigure}[b]{0.4\textwidth}
        \centering
\includegraphics[scale=.5]{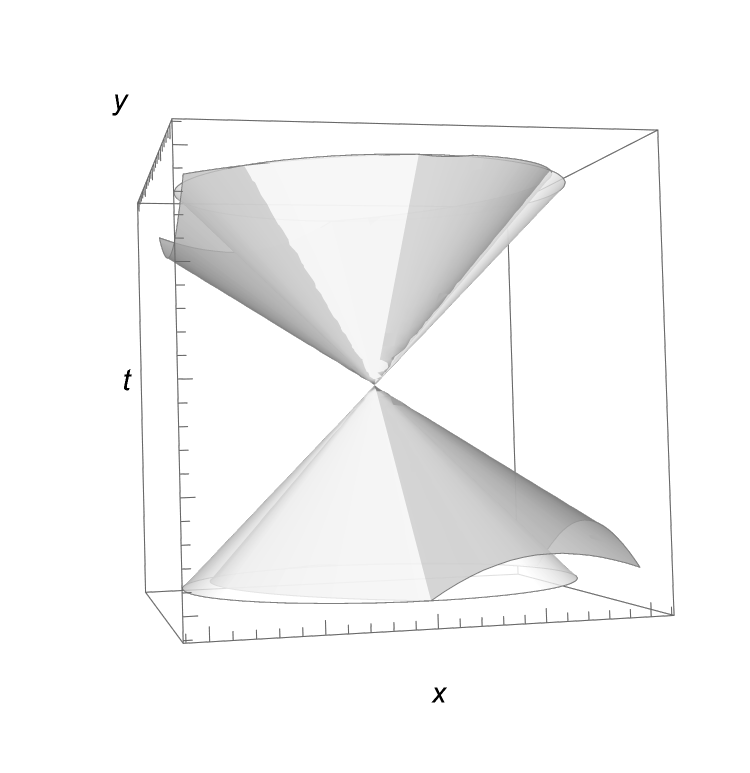}
\caption{A $t-x-y$ spacetime diagram illustrating a modified light cone (dark gray) and the standard light cone (light gray), which are the curves satisfying $t^2 - x^2 - y^2 + 2 k_{01}tx+k_{12}xy=0$ , 
and $t^2-x^2-y^2=0$, 
respectively}.
This shows the directional dependence of the modified symmetry-breaking case.
See also figure 1 in Ref. \cite{Schreck:2011ai}.
\label{fig2}  
\end{subfigure}
        \caption{Spacetime diagrams illustrating the modified propagation.}
        \label{fig:figs}
\end{figure}

We include here plots of how to visualize the propagation of the wave from the source point (taken as the origin) to the field point $t,x$; these can be seen in Figures~\ref{fig:figs}.  
Note the waves propagate in the direction $N_i=-\prt_i \tt_r$, with expanded form in \rf{N1}.
The coefficients used in the figure are the $k_{0i}$ coefficients, 
which are odd under Parity transformations.
This behavior is reflected in the first figure where the modified case breaks $x\rightarrow -x$ symmetry of the usual lightcone.

\subsection{Blending with perturbative solutions}

It is useful to compare the methods above with other methods that involve approximate solutions.
This has been carried out successfully in slow motion, weak field scenarios where wave behavior is not dealt with directly \cite{bk06,Bailey_2023}.
When the full effects of time derivatives is included, 
and we are looking for complete ``inhomogeneous" solutions 
(not just vacuum propagation), subtleties arise as we point out here.

To illustrate, 
we focus on the scalar wave equation case in \rf{scalar}.
The philosophy behind perturbative approaches is to seek solutions in powers of the small coefficients $k_\mn$.
For instance we assume
the solution can be written
$\ps=\ps^{(0)} + \ps^{(1)}+...$, 
with $(n)$ indicating order in powers of $k_\mn$.
To zeroth and first order in $k_\mn$, 
we have the two equations to solve,
\beq
\Box \ps^{(0)} = -\rh,
\quad
\Box \ps^{(1)} = -k^\mn \prt_\mu \prt_\nu \ps^{(0)}. 
\label{firstorder}
\eeq
The formal (particular) solutions, with the retarded time Green function, 
are
\beq
\ps^{(0)} =  
\int d^3 r^\prime \frac {\rh (t-R,\vec r^\prime ) }{R},
\quad
\ps^{(1)} = 
\int d^3 r^\prime \frac {k^\mn \prt^\prime_\mu \prt^\prime_\nu \ps^{(0)} }{R}
\label{pertsolns1}
\eeq
The first solution is the conventional scalar one, and
the second of these equations involves a field on the right-hand side 
that can be nonzero over all regions of space, 
and so does not have compact support.
Even in GR, 
integration of the formal wave solution also involves source terms 
composed of nonzero fields far from the source \cite{mtw,pw14}.
Such terms can be evaluated in GR 
and form part of the complete causal and properly behaved solution
\cite{pw00,pw14,Blanchet:2013haa}.
It is not immediately clear for equation \rf{firstorder}
if this program works.

We have solved equation \rf{pertsolns1} for the scalar example case of 
section \ref{exact solution} using standard methods \cite{pw14}.
The result truncates to leading order in the coefficients $k_\mn$ and the retarded time argument that appears is the standard one $t-r$, 
rather than the modified one in \rf{N1}.
The result is
\beq
\bal
\ps &= \frac {1}{4\pi r} \Big( 
Q [ 1- \tfrac 12 k_{00} + \tfrac 12 k_{ij}n^i n^j ]
+ \dot P^i 
[n_i (1- \tfrac 12 k_{00} + \tfrac 32 k_{jk} n^j n^k + k_{0j} n^j ) -k_{ij} n^j - k_{0i} ]
\\
&
\pt{space}
+\frac 12 \ddot I^{ij}
[n_i n_j ( 1 + 3 k_{lm}n^l n^m + 3 k_{0l} n^l ) 
-\de_{ij} ( \tfrac 12 k_{00} + k_{0l} n^l + \tfrac 12 k_{lm} n^l n^m ) 
\\
&
\pt{space space}
-2 n_i k_{jk} n^k - 2 k_{0i}n_j ]
+... \Big)|_{t=t_r}.
\eal
\label{psiLOpert}
\eeq
Comparison with \rf{psiLO} shows a mismatch of numerical factors and the absence of terms with the trace of $I^{ij}$.
It turns out that the two approaches indeed match but there is a subtlety that involves the correct conversion of the expression \rf{psiLO} from the modified retarded time $\tt_r$ to the usual $t_r = t-r$.
This confirmation suggests a general perturbative solution program exploring nonminimal gravity sector terms can be countenanced \cite{nilsson23}.

%Rui, please change to correct factors

Rather than solving \rf{pertsolns1} directly,
there is another alternative that more rapidly provides a match between perturbative approaches and ``exact" ones.
Applying the $\Box$ operator to the equation for $\ps^{(1)}$ in \rf{firstorder}, 
we obtain,
\beq
\Box^2 \ps^{(1)} 
= k^\mn \prt_\mu \prt_\nu \rh,
\label{alteqn}
\eeq
where now the right-hand side is a source with compact support but the left-hand side is a nonlocal operator.
A nonlocal Green function for the operator that solves
$\Box^2 G = -\de^4 (x-x^\prime)$ takes the form
$G_{nl}(x,x^\prime) = -(1/16\pi) {\rm sgn} (t-t^\prime \pm R)$,
where ${\rm sgn} (x)=\pm 1$: positive if $x>1$ and negative if $x<1$.
This result can be derived from standard sources, 
for example, 
by taking the Fourier time transform of the relevant position space 
Green functions in Ref.\ \cite{Lindell}.\footnote{The static limit of this Green function, 
which is just proportional to the distance $R$, 
is used ubiquitously in the literature for various post-Newtonian applications \cite{bk06,bh17,km17,Will:2018bme}.
Green functions for nonlocal operators have been discussed elsewhere, 
for instance Refs. \ \cite{Pais:1950za,Eliezer:1989cr,Bailey_2023}.}%add chandrasekhar cite
When derivatives are applied to $G(x,x^\prime)$, 
the light cone delta function emerges.  
For example, 
$\Box G_{nl}(x,x^\prime) = \de (t-t^\prime \pm R)/(4\pi R)$.

The solution to \rf{alteqn} then takes the form
\beq
\ps^{(1)} = -\int d^4 x^\prime 
G_{nl} (x,x^\prime ) 
k^\mn \prt^\prime_\mu \prt^\prime_\nu \rh^\prime
+\ps^{(1)}_H,
\label{nlsoln}
\eeq
where $\ps^{(1)}_H$ is a homogeneous solution satisfying $\Box^2 \ps^{(1)}_H=0$, 
and the prime on $\rh$ indicates dependence on the primed spacetime point $x^\prime$.
Convergence of the integrals for the infinite domain in \rf{nlsoln} depends on the source function $\rh$ asymptotic properties and the bounding surface of the four-dimensional integral. 
We assume $\rh$ is localized in space, 
vanishing outside some finite radius.
The time behavior is another matter.
One can always introduce a bounding surface, 
for example, 
the volume is the spacetime between two spacelike hypersurfaces at fixed values of time $t_2$ and $t_1$ (see figure 5.3a in Ref.\ \cite{mtw}).
Alternatively one can introduce an artificial exponential time falloff for the density $\rh \rightarrow \rh e^{-\ep |t|}$ to ensure the source vanishes as time approaches $\pm \infty$, 
as done in adiabatic switching.

Assuming that such a modification is applied to \rf{nlsoln}, 
so that it is finite, 
we proceed with integration by parts with the $\prt^\prime_\mu \prt_\nu^\prime$ derivatives.
The surface terms can either be eliminated by a choice of the homogeneous solution $\ps^{(1)}_H$ or they can be shown to vanish on the boundary with mild assumptions.
We obtain
\beq
\ps^{(1)} = -\int d^4 x^\prime 
k^\mn \prt^\prime_\mu \prt^\prime_\nu G_{nl} (x,x^\prime ) 
\rh^\prime,
\label{nlsoln2}
\eeq
and now the derivatives of the Green function $\sim {\rm sgn} (t-t^\prime \pm R)$
will always involve a delta function along the light cone.
The result in \rf{nlsoln2} is best matched to the ``exact" solution \rf{psi} 
by breaking up the summation into space and time components.
After evaluating derivatives using standard step and delta function properties,
and adding in the zeroth order solution, 
we collect the terms in a suggestive form:
\beq
\ps^{(0)}+\ps^{(1)} =  \int d^3 r^\prime  \frac {1}{4\pi R} \big[ \rh^\prime_r 
+ {\dot \rh}^\prime_r \frac 12 \big( k_{00} R  + 2 k_{0j} R^j + k_{jk} R^j {\hat R}^k\big)
+ \rh^\prime_r \frac 12 k_{jk} (\de^{jk} - {\hat R}^j {\hat R}^k ) \big],
\label{nlsoln4}
\eeq
where the subscript on $\rh$ indicates evaluation at the retarded time $t_R=t-R$.

The first term in \rf{nlsoln4} is the unperturbed solution for when $k_\mn=0$.
The second term has an unconventional dependence on the distance $R$;
far from the source the potential has no $1/r$ suppression.
However, 
the second term can be re-interpreted as the first order term in the Taylor expansion of time argument \rf{ttR}:
$\rh (\tt_R) = \rh (t_R)+ {\dot \rh} (\tt_R - t_R)+...$,
given that 
$\tt_R - t_R = \frac 12 ( k_{00} R  + 2 k_{0j} R^j + k_{jk} R^j {\hat R}^k ) + O(k^2)$.
The third term has the usual $1/r$ suppression outside the source region.
When comparing to the exact solution \rf{psi}, 
the third term can be understood as arising from a series expansion of
the modified distance ${\tilde R}$ in \rf{ttR};
$\tilde R = R (1 + \frac 12 k_{00} - \frac 12 k_{jk} {\hat R}^j {\hat R}^k + O(k^2))$.
To summarize we have shown the match of approximate and exact solutions:
\beq
\ps^{(0)}+\ps^{(1)} = \ps +O(k^2),
\label{match}
\eeq
but it should be noted that care was required to interpret apparent nonlocal terms.

\section{Photon sector application}
\label{photon sector application}

We apply the Green function formalism of section \ref{Green function}
to the photon sector of the EFT framework \cite{ck98,km02}.
The field equations from the photon sector action in the ``non-birefringence" limit can be written in the form,
\beq
\big( 
\et^{\mu\ka} \et^{\la\nu} +   
\et^{\mu\ka} (c_F)^{\la\nu} + 
(c_F)^{\mu\ka} \et^{\la\nu}
\big)
\prt_\mu F_{\ka\la} = -j^\nu,
\label{FEph}
\eeq
where $(c_F)^\mn$ are 9 coefficients for Lorentz violation (symmetric and assumed traceless),
$F_\mn = \prt_\mu A_\nu - \prt_\nu A_\mu$ is the field strength tensor, 
and $j^\nu$ is the current source \cite{km02,km09,Bailey:2010af}.
If we define an alternate set of coefficients $\Ct^\mn$ using
$\Ct^\mu_{\pt{\mu}\ka} ( \et^\ka_{\pt{\ka}\nu} + \tfrac 12  \Ct^\ka_{\pt{\ka}\nu} )= (c_F)^\mu_{\pt{\mu}\nu}$,
then we can write \rf{FEph} as 
\beq
\tg^{\mu\ka} \tg^{\la\nu} 
\prt_\mu F_{\ka\la} = -j^\nu,
\label{FEph2}
\eeq
where $\tg^\mn = \et^\mn + \Ct^\mn$, 
similar to the definition in \rf{modMetric}.
Note that to leading order in small dimensionless coefficients, $\Ct^\mu_{\pt{\mu}\ka} \approx (c_F)^\mu_{\pt{\mu}\ka}$.

To solve this equation, 
we change coordinates $x^\mu = x^\mu ( \xb^\nu ) $, 
in the same manner as \rf{gchange}.
In the new coordinate system the field equations take the form
\beq
\et^{\mu\ka} \et^{\la\nu} 
\oprt_\mu \oF_{\ka\la} = -{\overline j}^\nu,
\label{FEph3}
\eeq
with $\oprt_\mu= \prt/\prt \xb^\mu$ and $\oF_\mn = \oprt_\mu \oA_\nu - \oprt_\nu \oA_\mu$.
The field equations resemble that of conventional electrodynamics
in the $\xb^\mu$ coordinates
with a modified current $j^\nu$.
In particular,
there remains the usual gauge symmetry of electrodynamics.
We adopt the gauge choice
$\et^\mn \oprt_\mu \oA_\nu=0$, 
leaving the field equations as
\beq
\et^\mn \oprt_\mu \oprt_\nu \oA_\la = -{\overline j}^\nu \et_{\nu\la},
\label{xbFEgf}
\eeq

where we have multiplied \rf{FEph3} by a Minkowski inverse $\et_\mn$ on both sides to isolate $\oA_\la$.
The standard wave operator appears in equation \rf{xbFEgf} and so the usual inhomogeneous solution can be used, 
yielding
\beq
\oA_\la = \frac {1}{2\pi}  \int d^4 \xb^\prime
\de \big( -\et_{\al\be} (\xb -\xb^\prime)^\al (\xb -\xb^\prime)^\be \big)
{\overline j}^{\prime \mu} \et_{\mu\la}.
\label{Abarsoln}
\eeq

Now we use the coordinate transformation rule
$A_\la = (\prt \xb^\mu / \prt x^\la) \oA_\mu$
and change the coordinates within the integral in \rf{Abarsoln}.
First, using equation \rf{gchange}, we can show the argument of the delta function in the original $x^\mu$ coordinates takes the form like \rf{GreenResult},
namely $-(\tg^{-1})_\mn (x-x^\prime )^\mu (x-x^\prime )^\nu$.
The remainder of the transformation follows from standard formulas.
The Jacobian of the transformation can be found from \rf{gchange} and can be written $|\prt \xb / \prt x|=1/\sqrt{-\tg}$.
The originally sought solution is then
\beq
A_\la = 
\frac {1} {2\pi \sqrt{-\tg }} 
\int d^4x^\prime 
\de \left( -(\tg^{-1})_{\al\be} (x-x^\prime )^\al (x-x^\prime )^\be \right)
j^{\prime \mu} (\tg^{-1})_{\mu\la}.
\label{Asoln}
\eeq
This result can be independently checked by using equation \rf{GreenEqn} and \rf{GreenResult}, 
and using the gauge condition transformed to the original coordinates,
namely $\tg^\mn \prt_\mu A_\nu=0$, 
to show that \rf{Asoln} satisfies \rf{FEph2} and hence
solves \rf{FEph}.

Following steps similar to those for the scalar field we can write the solution compactly as
\beq
A_\la = \frac {1} {4\pi \sqrt{-\tg }} 
\int d^3x^\prime 
 \frac {(\tg^{-1})_{\mu\la} j^\mu (\tt_R, \vec r^\prime)}{\tR},
\label{Asoln2}
\eeq
with $\tt_R$ and $\tR$ as in equations \rf{ttR}.
We specialize \rf{Asoln2} to a localized conserved current density 
$j^\mu = (\rh,\vec J)$
and expand the solution assuming the field point is far from the source and in the wave zone ($r \gg \la  \gg r^\prime$).
We follow steps similar to those leading up to
\rf{psiWZ}
and we arrive at
\beq
A_\la = \frac {(\tg^{-1})_{\mu\la}}{4\pi \sqrt{-\tg} }
\sum_{l=0}^{\infty} 
\frac {(-1)^l} {l!}
\prt_L \left( \frac {1}{\tr} 
\int d^3 r^\prime j^\mu (\tt_r , \vec r^\prime ) r^{\prime L} \right),
\label{AWZ}
\eeq
where $\tt_r$ and $\tr$ are defined in \rf{N1}.
We write out the terms up to $L=1$, 
decomposing the current into charge density and current density, 
and keeping only terms that fall off as $1/\tr$, 
to obtain
\beq
A_\mu = \frac {1}{4\pi \sqrt{-\tg} \tr } 
\int d^3r^\prime \Big[
(\tg^{-1})_{\mu 0} \rh^\prime  
+(\tg^{-1})_{\mu i} J^{\prime i} 
+ N_i (\tg^{-1})_{\mu 0} \dot {\rh^\prime} r^{\prime i} 
+...\Big]|_{t=\tt_r}.
\label{AWZ2}
\eeq
The first term is proportional to the constant total charge $Q$, while the second and third term can be re-expressed in terms of the electric dipole moment
$p^j = \int d^3r^\prime \rh^\prime r^{\prime j}$, 
using standard techniques 
\cite{Jackson}.
The higher order terms contribute to the magnetic dipole and quadrupole terms, 
which we neglect here.
The dominant {\it radiation} four-potential terms are 
\beq
A_\mu = \frac {1}{4\pi \sqrt{-\tg} \tr } 
\Big[ (\tg^{-1})_{\mu i} + (\tg^{-1})_{\mu 0} N_i \Big] {\dot p}^i|_{t=\tt_r}.
\label{radiation}
\eeq

The radiation zone electric field, which is gauge independent, 
is found to be
\beq
F_{i0} = -\frac {1}{4\pi \sqrt{-\tg} \tr } 
\Big[ (\tg^{-1})_{0 i} N_j + (\tg^{-1})_{0 j} N_i 
+(\tg^{-1})_{ij} 
+ (\tg^{-1})_{00}N_i N_j
\Big] {\ddot p}^j|_{t=\tt_r}.
\label{electric}
\eeq
Adopting a leading order expansion with $(\tg^{-1})_\mn = \et_\mn - (c_F)_\mn$ and using results above such as \rf{N1}, 
we can also write the electric field as
\beq
F_{i0} = -\frac {1}{4\pi \sqrt{-\tg} \tr } 
\Big[ P_{ij} + P_{ik} P_{jl} (c_F)^{kl} \Big] {\ddot p}^j|_{t=\tt_r},
\label{electricLO}%leading order
\eeq
where $P_{ij} = \de_{ij} - n_i n_j$
is a projection operator.
Note that while $F_{i0}n_i=0$, 
indicating two independent polarizations, 
$n_i$ is not the true direction of wave propagation.
There remains a projection along the true direction of propagation $N_i$ that is not zero, 
$F_{0i} N_i \neq 0$, 
indicating that at least one of the two independent modes 
is not transverse to the wave propagation direction.

This striking result does not mean the usual $U(1)$ gauge invariance is broken. The classical result we obtained for electric field in \rf{electric} is gauge invariant, 
any piece of the vector potential $A_\mu$ proportional to $\partial_\mu$ vanishes in this calculation.  
However, 
with Lorentz violation in the form of the $c_F$ coefficients, 
the electric field can have a longitudinal component even in the vacuum.
From the modified Maxwell equations in the photon sector, 
we get $\vec \nabla \cdot \vec E = - \vec \nabla \cdot \kappa_{DE} \cdot  \vec E - \vec \nabla \cdot \kappa_{DB} \cdot  \vec B $, 
unlike the usual Maxwell vacuum case where $\vec \nabla \cdot \vec E=0$ (using the matrix notation for the coefficients $\kappa_{DE}$ and $\kappa_{DB}$ of Ref.\ \cite{km02}).

\section{Gravity sector application}

\subsection{General solution}
\label{general solution}

For a starting point, 
we use the EFT gravity sector field equations for the metric fluctuations
$h_\mn$ around a flat background.
They can be obtained from a Lagrange density 
$\cL = \cL_{GR} + \tfrac {1}{4\ka} \sb^{\al\be} h^\mn \cG_{\al\mu\be\nu}+\cL_M$,
where $\cG_{\al\mu\be\nu}$ is the double dual of the Riemann tensor 
and $\ka=8 \pi G_N$ \cite{Bailey:2013fwa,km16,bkx15,bh17}.
We can write the field equations in the form
\beq
{\hat K}^{\mu\nu\al\be} h_{\al\be} = \ka \ta^\mn,
\label{FE1}
\eeq
where $\ta^\mn$ includes the matter stress energy tensor $(T_M)^\mn$
as well as contributions from higher order terms in $h_\mn$ 
with and without coefficients for Lorentz violation.
Should the coefficients $\sb_\mn$ arise dynamically, 
through spontaneous symmetry breaking, 
the dynamical terms contributing to $\ta^\mn$ can also be included
\cite{b21}.
The operator ${\hat K}^{\mu\nu\al\be}$ can be written
as
\beq
\bal
{\hat K}^{\mu\nu\al\be} &= \frac 12 \Big( 
\et^{\al (\mu} \et^{\nu )\be} \et^\gd 
-\et^\mn \et^\ab \et^\gd 
+\et^\mn \et^{\al\ga} \et^{\be\de}
+ \et^\ab \et^{\mu\ga} \et^{\nu\de}
-\et^{\al (\mu} \et^{\nu ) \ga} \et^{\be\de}
-\et^{\be (\mu} \et^{\nu ) \ga} \et^{\al\de}
\Big) \prt_\ga \prt_\de
\\
&
+
{\hat K}_s^{\mu\nu\al\be},
\eal
\label{Khat}
\eeq
where ${\hat K}_s^{\mu\nu\al\be}$ is the operator such that 
${\hat K}_s^{\mu\nu\al\be} h_{\al\be} = \sb_{\al\be} {\cal G}^{\mu\al\be\nu}$.
The first line in \rf{Khat} contains the terms present in standard linearized GR, 
namely the terms in $G^\mn$, while ${\hat K}_s^{\mu\nu\al\be} h_{\al\be}$ is the leading order corrections 
from the $\sb^\mn$ coefficients \cite{bk06,km16}.

For the purposes in this work, 
it is useful to re-express the operator \rf{Khat} in a simpler form.
We define $\tg^\mn = \et^\mn + \sb^\mn$. 
Then to first order in $\sb^\mn$ it can be shown that
\beq
{\hat K}^{\mu\nu\al\be} = \frac 12 \Big( 
\tg^{\al (\mu} \tg^{\nu )\be} \tg^\gd 
-\tg^\mn \tg^\ab \tg^\gd 
+\tg^\mn \tg^{\al\ga} \tg^{\be\de}
+ \tg^\ab \tg^{\mu\ga} \tg^{\nu\de}
-\tg^{\al (\mu} \tg^{\nu ) \ga} \tg^{\be\de}
-\tg^{\be (\mu} \tg^{\nu ) \ga} \tg^{\al\de}
\Big) \prt_\ga \prt_\de ,
\label{Khat2}
\eeq
which resembles the standard linearized terms in GR but with an apparent modified background metric $\tg^\mn$, 
as pointed out in \cite{kt11}.

We perform a general coordinate transformation as in the scalar and vector case above.
We require the coordinate transformation to satisfy \rf{gchange}, 
with $\tg^\mn = \et^\mn + \sb^\mn$.
Treating the quantities in \rf{Khat2} as tensors in a flat background, 
the field equations in the 
${\overline x}^\mu$ coordinates take the form
\beq
\frac 12 \Big( 
\et^{\al (\mu} \et^{\nu )\be} \et^\gd 
-\et^\mn \et^\ab \et^\gd 
+\et^\mn \et^{\al\ga} \et^{\be\de}
+ \et^\ab \et^{\mu\ga} \et^{\nu\de}
-\et^{\al (\mu} \et^{\nu ) \ga} \et^{\be\de}
-\et^{\be (\mu} \et^{\nu ) \ga} \et^{\al\de}
\Big) 
{\overline \prt}_\ga {\overline \prt}_\de \oh_{\al\be} = \ka {\overline \ta}^\mn .
\label{xbarFE}
\eeq
Thus in this coordinate system, 
the field equations appear as conventional linearized GR with a modified source 
${\overline \ta}^\mn $.\footnote{The barred notation indicates the coordinate system 
and is not to be confused with the common trace-reversed bar notation.}

Next we exploit the gauge freedom in equation \rf{xbarFE} and choose
\beq
\et^{\al\be} {\overline \prt}_\al \oh_{\be\ga} = \frac 12 {\overline \prt}_\ga ( \et^{\al\be} \oh_{\al\be}).
\label{GravGauge}
\eeq
Note that ${\overline \prt}_\la \et^\mn =0$ holds in the $\xb^\mu$ coordinates.
With the gauge choice, 
the field equations become
\beq
\frac 12 \et^\ab {\overline \prt}_\al {\overline \prt}_\be \Pi^\mn = -\ka {\overline \ta}^\mn, 
\quad 
{\rm with} \quad
\Pi^\mn = \et^{\mu\al} \et^{\nu\be} \oh_\ab - \frac 12 \et^\mn ( \et^{\al\be} \oh_{\al\be}).
\label{gfFEbar}
\eeq

The standard wave operator (in ${\xb^\mu}$ coordinates) appears in \rf{gfFEbar}, 
thus we can use the standard wave solution,
\beq
\Pi^\mn  = \frac {\ka}{\pi} \int d^4 \xb^\prime \de \big( -\et_{\al\be} (\xb -\xb^\prime)^\al (\xb -\xb^\prime)^\be \big) {\overline \ta}^\mn.
\label{PiSoln}
\eeq
Using the Minkowski metric $\et_\mn$ we obtain $\oh_\mn$ from \rf{PiSoln}:
\beq
\oh_\mn = \frac {\ka}{\pi} \int d^4 \xb^\prime \de \big( -\et_{\al\be} (\xb -\xb^\prime)^\al (\xb -\xb^\prime)^\be \big)  
\left( \et_{\ga\mu} \et_{\de\nu} {\overline \ta}^{\ga\de}  - 
\frac 12 \et_\mn \et_\gd {\overline \ta}^\gd \right).
\label{ohSoln}
\eeq

Using the coordinate transformation rule
$h_{\ka\la} = (\prt \xb^\mu /\prt x^\ka) (\prt \xb^\nu /\prt x^\la) \oh_\mn $,
we can find the solution in the original coordinates, 
similar to the approach for the vector potential in the steps leading to \rf{Asoln}.
This yields
\beq
\bal
h_\mn &= 
\frac {\ka} {\pi \sqrt{-\tg }} 
\int d^4x^\prime 
\de \left( -(\tg^{-1})_{\al\be} (x-x^\prime )^\al (x-x^\prime )^\be \right)
\\
&
\pt{space}
\times
\left( 
(\tg^{-1})_{\ga\mu} (\tg^{-1})_{\de\nu} \ta^\gd  
-\frac 12 (\tg^{-1})_\mn (\tg^{-1})_\gd \ta^\gd 
\right).
\label{hSoln}
\eal
\eeq
We have also directly verified that this solution \rf{hSoln} solves equation \rf{FE1} 
to leading order in $\sb_\mn$.  
Note that the gravitational wave from the source propagates along the modified light cone as in section \ref{scalar example}, 
which is consistent with prior propagation studies \cite{km16,Mewes:2019,ONeal-Ault:2021uwu}.
What is new here is that we can calculate directly the effects of a given source
on the metric fluctuations and the measured effects in a GW detector.

In a leading order approximation, 
we have $(\tg^{-1})_\mn =\et_\mn -\sb_\mn$.
If we expand the delta function to integrate over $t^\prime$, 
as done for the scalar case and vector case above, 
and we restrict attention to leading order in $\sb_\mn$, 
then we obtain the result
\beq
h_\mn = 
\frac {\ka} {2\pi \sqrt{-\tg }} 
\int d^3x^\prime 
\frac {1}{\tR} \Big( 
\ta_\mn - 2 \ta^\al_{\pt{\al}(\mu} \sb_{\nu )\al} 
-\frac 12 \et_\mn (\ta^\al_{\pt{\al}\al}  - \sb_\ab \ta^\ab ) +\frac 12 \sb_\mn \ta^\al_{\pt{\al}\al}
\Big) (\tt_R, \vec r^\prime),
\label{hLO}
\eeq
where $\tt_R$ and $\tR$ are defined in \rf{ttR}, 
with $k^\mn \rightarrow \sb^\mn$.
This solution is valid in the gauge
\beq
(\et^\mn + \sb^\mn ) \prt_\mu h_{\nu\la} =
\frac 12 \prt_\la 
( \et^\mn + \sb^\mn ) h_{\mn},
\label{gauge2}
\eeq
which is not the usual harmonic gauge unless $\sb_\mn=0$ \cite{Xu:2019fyt}.

\subsection{Expansion of solution}
\label{expansion}

At this stage we employ the far field expansion, 
similar to \rf{AWZ}.
First we abbreviate the terms in parenthesis inside the integral \rf{hLO} as $\Th_\mn$.
We seek the solution for $h_\mn$ in the far field or wave zone.
However, 
we must integrate over the
near zone $\cN$ and wave zone $\cW$ in this case because $\ta_\mn$ does not have compact support and exists in both regions:
\beq
h_\mn = \frac {4G} { \sqrt{-\tg } } \Big(  %2 pi removed 
\int_\cN d^3x^\prime \frac {\Th_\mn (\tt_R, \vec r^\prime)}{\tR}+
\int_\cW d^3x^\prime \frac {\Th_\mn (\tt_R, \vec r^\prime)}{\tR}
\Big).
\label{NW}
\eeq
The integrals over the wave zone involve those contributions 
to $\Th_\mn$ that do not have compact support; 
they are of higher order in a series in powers of $h_\mn$ (or equivalently powers of $G$ \cite{pw14}).  
As this paper is more of an introductory nature, 
we attempt only the first integrals, 
so we seek $(h_\cN)_\mn$, 
and leave the calculation of $(h_\cW)_\mn$ for future work.

The general solution for the $\cN$ zone integrals can be put into an expansion form
like \rf{psiWZ}:
\beq
(h_\mathcal{N})_\mn = \frac {4G} {\sqrt{-\tg }}
\sum_{l=0}^{\infty} 
\frac {(-1)^l} {l!}
\prt_L \left( \frac {1}{\tr} 
\int_\cN d^3 r^\prime \Th_\mn (\tt_r, \vec r^\prime) r^{\prime L} \right).
\label{GWN}
\eeq
We proceed to evaluate the first few terms in the series \rf{GWN}
in order to find the leading multipole terms up to the quadrupole,
the latter being the traceless version of the inertia tensor $I^{ij}$:
\beq
I^{ij} = \int d^3 r \ta^{00} r^i r^j.
\label{inertia}
\eeq
Integrals in \rf{GWN} involve the space and time projected components of $\ta^\mn$; namely $\ta^{00}$, $\ta^{0j}$, and $\ta^{jk}$.  
The goal for a concise solution is to express all the terms using the inertia tensor \rf{inertia}.
We can use the conservation law $\prt_\mu \ta^\mn = 0$, 
to express some of the integrals in \rf{GWN} in terms of $\int d^3 r \ta^{ij}$; this quantity can be re-expressed in terms of the inertia tensor. 
The latter step is achieved with the identity $\prt_0^2 \ta^{00} =\prt_i \prt_j \ta^{ij}$ \cite{pw14}:
\beq
\int d^3 r \ta^{ij} =(1/2) d^2/dt^2 \int d^3 r \ta^{00} r^i r^j + \partial \cN {\rm terms}.
\label{identity}
\eeq
The boundary terms are dependent on $\cR$ and are expected to cancel with corresponding terms from the wave zone integrals.

For the radiation fields $(h_\cN)_\mn$ we find, 
up to surface terms at radius ${\cal R}$,
\beq
\bal
(h_\cN)_{00} &= \frac {G}{\tr} 
\big( \de_{jk} + N_j N_k 
+\sb_{00} (\de_{jk} + 2 N_j N_k) 
%\pt{spa}
+2 \sb_{0j} N_k -\sb_{jk}\big) \ddot{I}^{jk}|_{t=\tt_r},
\\
(h_\cN)_{0j} &= \frac {G}{\tr} \big( -2 \de_{jk} N_l (1+\sb_{00})  -2 \sb_{0k}\de_{jl} + \sb_{0j}\de_{kl}
%\pt{spa}+
+\sb_{0j} N_k N_l + 2 \sb_{jk} N_l 
\big) \ddot{I}^{kl}|_{t=\tt_r},
\\
(h_\cN)_{jk} &= \frac {G}{\tr} 
\big(  
2 \de_{l(j} \de_{k)m} - \de_{jk} \de_{lm} 
+\de_{jk} ( 1+\sb_{00} ) N_l N_m 
- 4 \sb_{l(j} \de_{k)m} 
- \sb_{jk} N_l N_m 
\\
&
-4 \sb_{0(j} \de_{k)m} N_l 
+ 2 \sb_{0m}\de_{jk} N_l 
+\de_{jk} \sb_{lm} +\de_{lm} \sb_{jk} \big) 
\ddot{I}^{lm}|_{t=\tt_r},
\label{componentsWZ}
\eal
\eeq
Since the focus is on the radiation fields, 
we omit the near zone potentials which can be found in Ref.\ \cite{bk06}.
The measured curvature in a gravitational wave detector 
can be taken as the components 
$R_{0j0k}=(1/2)(\prt_0 \prt_j h_{0k}+\prt_0 \prt_k h_{0j} - \prt_j \prt_k h_{00} - \prt^2_0 h_{jk})$ \cite{mtw}.  
Normally in GR, in the usual transverse traceless gauge, 
one can obtain the curvature directly from $h_{jk}$ alone.
The gauge choice made here does not generally allow that;  
however, 
the curvature is gauge independent,
hence our focus on observable effects.
We find the curvature components to be 
\beq
\bal
R_{0j0k} &= \frac {G}{\tr} 
\big[ 
\tfrac 12 \de_{jk} \de_{lm} - \de_{l(j} \de_{k)m} 
-\tfrac 12 \big(\de_{jk}  N_l N_m 
+\de_{lm} N_j N_k 
-4 \de_{l (j} N_{k)} N_m \big) 
(1+ \sb_{00})
\\&
- \tfrac 12 N_j N_k N_l N_m (1+2 \sb_{00})
+2\sb_{0(j} \de_{k)m} N_l
+ \sb_{0m} \de_{jk} N_l 
 +2 \sb_{0m} \de_{l(j} N_{k)} 
\\&
-\sb_{0(j} N_{k)}\de_{lm}
- \sb_{0m} N_j N_k N_l 
- \sb_{0(j} N_{k)} N_l N_m
-\tfrac 12 \big(  \de_{jk} \sb_{lm} 
+ \de_{lm} \sb_{jk} 
\\&
- 4 \sb_{l(j} \de_{k)m} - \sb_{jk} N_l N_m 
%\pt{+\tfrac 12}
- \sb_{lm} N_j N_k
+4 \sb_{l(j } N_{k)} N_m 
\big) \big] (\overset{(4)}{I})^{lm}|_{t=\tt_r}
\label{curvature}
\eal
\eeq
Henceforth all time-dependent quantities will be evaluated at $t=\tt_r$ and we omit the $|t=\tt_r$ notation.

In general metric models of gravity beyond GR, 
there are up to six possible polarizations for gravitational waves \cite{Will:2018bme,Schumacher:2023jxq}.
In the presence of Lorentz violation 
in \rf{curvature},
five of the six polarizations show up.
We can also establish the question of their independence, 
and the number of degrees of freedom.
We will identify the polarizations by taking the trace and projections
of $R_{0i0j}$.
Since the wave travels in a direction along $N_i$ we will adopt a spatial basis
$\{ {\bf e}_1, {\bf e}_2, \bf N/\sqrt{N^i N_i} \}$, 
where the basis vectors ${\bf e}_1$ and ${\bf e}_2$ span the plane perpendicular to $N_i$.
Note that, 
due to the coefficients in \rf{N1}, 
${\bf e}_1$ and ${\bf e}_2$ are not perpendicular to $n^i$,
except at zeroth order in the coefficients.

First we calculate the trace of the curvature tensor $R_{0j0k}\de^{jk}$.
It will be convenient to introduce a traceless $(\sb_{tr})_{ij}=\sb_{ij} - (1/2)\de_{ij} \sb_{00}$, 
where we use the assumption $\sb^\mu_{\pt{\mu}\mu} = \sb_{jj} - \sb_{00}=0$.
The trace can be simplified to
\beq
R_{0\pt{j}0j}^{\pt{0}j} = \frac {G}{\tr} \left[ (\sb_{tr})_{\perp ij}+ \frac 12 (\sb_{tr})_{nn} (\de_{ij} - n_i n_j ) \right] 
(\overset{(4)}{I})^{ij},
\label{trace}
\eeq
where projections of quantities along $\hat n$ are denoted with the index $n$ 
and $\perp$ indicates a projection of a tensor perpendicular to $\hat n$ like
$(V_\perp)_i = V_i - n_i V_j n^j = V_i - n_i V_n$. Note that to leading order in the coefficients $\sb_\mn$ we replace $N^i$ with $n^i$ in \rf{trace} and elsewhere below.

Next we find the double projection of the curvature along the wave propagation direction $N^i N^j R_{0i0j}$.  
We find
\beq
N^i N^j R_{0i0j}=0+O(\sb^2),
\label{nnR}
\eeq
thus there is no leading order polarization along this projection.
We can then conclude then that the scalar projection onto the transverse plane, 
$(\de^{ij} - N^i N^j)R_{0i0j}=R_{0101}+R_{0202}$, 
is the same as the trace in \rf{trace}.
However, 
the components $R_{0i0j} N^i (e_a)^j$ do not vanish (where $a=1,2$).
They are given by
\beq
R_{0i0j} N^i (e_a)^j = \frac {G}{\tr} \big[ \tfrac 12 \big( (\sb_{tr})_{an}+ \sb_{0a} \big)  (\de_{ij} - n_i n_j ) 
+(e_a)_j \big( (\sb_{tr})_{nk_{\perp}} + \sb_{0k_\perp } \big)  \big] (\overset{(4)}{I})^{ij}.
\label{neproj}
\eeq

Finally, 
we display projections along the transverse directions ${\bf e}_1$ and ${\bf e}_2$, 
the ones that normally are called ``plus" and ``cross".
They are given by 
\beq
\bal
R_{0202} - R_{0101}& 
=
\frac {G}{\tr} \big[ 
(e_{1i} e_{1j} - e_{2i} e_{2j})(1-\tfrac 23 \sb_{00})
%\pt{space}
%\pt{space}
-2 ( (\sb_{tr})_{1i} e_{1j} -(\sb_{tr})_{2i} e_{2j} )
\\&
%\pt{space}
-2 ( \sb_{01} e_{1i} n_j - \sb_{02} e_{2i} n_j )
+ \tfrac 12 ( (\sb_{tr})_{11} - (\sb_{tr})_{22} ) (\de_{ij} - n_i n_j)
\big] 
(\overset{(4)}{I})^{ij},
\\
R_{0102} &= 
\frac {G}{\tr} \big[
- (e_1)_i (e_2)_j (1-\tfrac 23 \sb_{00} ) 
+(\sb_{tr})_{1i} (e_2)_j
%\pt{space}
+(\sb_{tr})_{2i} (e_1)_j
\\&
-\tfrac 12 (\sb_{tr})_{12}
(\de_{ij} - n_i n_j)
%\pt{space}
+\sb_{01} (e_2)_i n_j + \sb_{02} (e_1)_i n_j \big] (\overset{(4)}{I})^{ij}
\label{pluscross}
\eal
\eeq
where the subscripts $1$ and $2$ imply projection with the corresponding unit vectors.
It should be noted that the results in \rf{pluscross} could also receive $\sb_\mn$ terms from the inertia tensor $I^{ij}$ itself.
Such terms could arise due to orbital effects from $\sb_\mn$ on a binary source, for example \cite{bk06}.
A self-gravitating system was shown to be affected in this manner \cite{Xu:2019gua}.
For brevity, a study of these effects is omitted here.

In GR, 
all projections but $R_{0202}-R_{0101}$ and $R_{0102}$ vanish (when $\hat n$ is the $3$ direction), 
as can be seen by setting all $\sb_\mn$ coefficients to zero.
In the presence of the coefficients it appears $3$ additional polarizations arise.
The results above indicate that the coefficients $\sb_\mn$, 
in addition to showing up in weak-field gravity scenarios like solar system tests \cite{Hees:2016lyw},
and affecting the speed of gravitational waves \cite{km16},
can also affect the observed polarization content in a GW detector.
The additional polarizations are of order $\sb$.  
Given the sensitivity of the current detectors to the strength of the GW signals above noise level of a couple orders of magnitude, 
it seems that these additional effects could be observed if $\sb \sim 10^{-2}$.

Constraints on all nine coefficients $\sb_\mn$ already exist below parts in $10$ billion (e.g., from lunar laser ranging \cite{Bourgoin:2016ynf}), 
so we do not expect observable effects in GW measurements via searches for extra polarizations.
However, 
we have not studied the effects of higher-order terms in the action \cite{km16}, 
and many of these coefficients are not well constrained, 
or not constrained at all, 
so are the subject of future work.

While we do not discuss details, 
the nonzero projections found are equivalent to some of the Newman-Penrose projections of the curvature tensor \cite{np62, Will:2018bme}.
Specifically we have 
\beq
\bal
& (\de^{ij} - N^i N^j)R_{0i0j} =-2\Ph_{22}, \quad
&N^i N^j R_{0i0j} =-6 \Ps_2 = 0, 
\\
&R_{0i0j} N^i (e_1)^j =-2\sqrt{2} {\it Re} {\Ps_3}, \quad
&R_{0i0j} N^i (e_2)^j =2\sqrt{2} {\it Im} {\Ps_3},
\\
&R_{0202}-R_{0101} = 2 {\it Re} {\Ps_4}, \quad
&R_{0102} = {\it Im} {\Ps_4}.
\label{np}
\eal
\eeq
The reader can refer to depictions of the effect of these modes on a sphere of test masses in Refs.\ \cite{Will:2018bme,Wagle:2019mdq}.

Finally, 
we comment regarding the number of independent degrees of freedom indicated by the five curvature polarizations.
It can be shown that three of them, 
the beyond-GR projections in \rf{trace} and \rf{neproj} can be written as linear combinations of the ``plus" and ``cross" polarizations in \rf{pluscross}.
This holds to first order in the coefficients, $\sb_\mn$.
Therefore we can say that at leading order in small Lorentz violation, only two propagating degrees of freedom remain, 
which is consistent with other results \cite{km18,Liang:2022hxd}.

\section{Summary}

In this letter we found the classical radiation fields for modified wave equations that occur in descriptions of spacetime-symmetry breaking.
The main results of the paper include the generic Green function solution \rf{GreenResult}, which can be applied to several cases.
In the presence of minimal forms of Lorentz violation, 
we found the general solution for retarded boundary conditions for a scalar field \rf{psi}, 
the vector potential \rf{Asoln2}, 
and the metric fluctuations \rf{hLO}.
These results were studied in a radiation zone expansion, 
with scalar results in \rf{psiLO}, 
the modified dipolar electric field \rf{electric},
and spacetime curvature from a gravitational wave source \rf{curvature}.
We found that Lorentz violation modifies the electric field so that the two independent components of the radiation fields from an electric dipole are not transverse to the direction of wave propagation, unlike in conventional electrodynamics.  The latter effect persists despite the theory maintaining the usual $U(1)$ gauge invariance.
For the partial solution we obtained for gravitational wave generation, 
there are 3 extra polarizations beyond GR. These polarizations are linear combinations of the plus and cross polarizations; thus overall there are still only two propagating degrees of freedom.

Results can be further studied in various ways.
For gravitational waves,
one needs a complete evaluation of \rf{hLO} including the contributions from the wave zone integrals $(h_\cW)_\mn$.
Note that we have not considered in detail the effects of the Nambu-Goldstone and massive modes that may occur from a spontaneous symmetry breaking scenario \cite{ks89bb,bk05,bk08}.
A general description of the dynamical terms for the $\sb_\mn$ coefficients, 
when they arise as a vacuum expectation value of a dynamical tensor $s_\mn$
has been published, 
but not yet studied in the GW context \cite{b21}.
A study of the multipole radiation expansion results in section \rf{photon sector application} could be carried out, 
for example looking for new possible observables for Lorentz violation in experiments and observation complementing prior work \cite{km02}.
Results can also be extended to the nonminimal terms in the EFT framework \cite{nilsson23}.
Symmetry-breaking terms in the action more recently countenanced could also be of interest \cite{kl21}.

\section{Acknowledgments}
We wish to thank Jay D.\ Tasson for valuable comments and discussion.
Q.G.B.\ and A.S.G.\ were supported by the National Science Foundation under grants number 2207734 and 2308602.
N.A.N. was financed by CNES and acknowledges support by PSL/Observatoire de Paris. R.X.\ and L.S. were supported by the National Natural Science Foundation of China (11975027, 11991053), the
National SKA Program of China (2020SKA0120300), and the Max Planck Partner Group
Program funded by the Max Planck Society.

\bibliographystyle{elsarticle-num}
\bibliography{refs}

\end{document}